\documentclass[fleqn,usenatbib]{mnras}
\usepackage{hyperref}
\usepackage[T1]{fontenc}
\usepackage{ae,aecompl}
\usepackage{amsmath}
\let\oldAA\AA
\renewcommand{\AA}{\text{\normalfont\oldAA}}
\usepackage{array}
\usepackage{makecell}
\usepackage{amssymb}
\usepackage{booktabs}
\usepackage{pdflscape} 
\usepackage{footnote}
\usepackage{threeparttable}
\usepackage{epsfig}
\usepackage{pslatex}
\usepackage{subfig}
\usepackage{comment}
\usepackage{enumerate}
\usepackage{epstopdf}
\usepackage{csquotes}
\usepackage{multirow}
\usepackage{orcidlink}
\usepackage{soul}
\usepackage{multicol}
\usepackage{graphicx}
\usepackage{pdflscape}
\usepackage{booktabs}
\DeclareRobustCommand{\VAN}[3]{#2}
\let\VANthebibliography\thebibliography
\def\thebibliography{\DeclareRobustCommand{\VAN}[3]{##3}\VANthebibliography}

\graphicspath{{./}{figures/}}

\newcommand{\kms}{\,km\,s$^{-1}$}
\newcommand{\ergs}{\,erg\,s$^{-1}$}

\title[RS Oph 2006-2021]{Spectroscopic study of the Quiescent Stages in between the 2006 and 2021 outbursts of RS Ophiuchi
}
\author[]{
Gesesew R. Habtie\,\orcidlink{0000-0001-9827-738X}$^{1,2}$\thanks{E-mail: meetgesese@gmail.com},
Ramkrishna Das\,\orcidlink{0000-0002-5440-7186}$^{1}$
\\
$^{1}$ S N Bose National Centre for Basic Sciences, Salt Lake, Kolkata 700 106, India\\
$^{2}$ Debre Berhan University, Debre Berhan, Ethiopia\\ 
}

\date{Accepted 2025 January 17. Received 2025 January 17; in original form 2024 July 08 }

\begin{document}
\label{firstpage}
\pagerange{\pageref{firstpage}--\pageref{lastpage}}
\maketitle

\begin{abstract}
This paper presents a comprehensive spectroscopic analysis of the quiescent stage of the recurrent nova RS Ophiuchi between its 2006 and 2021 outbursts. The spectra shows prominent low-ionization emission features, including hydrogen, helium, iron emissions, and TiO absorption features. The \ion{H}{$\alpha$} and \ion{H}{$\beta$} lines showed double-peaked emission profiles, indicating that both originate from the accretion disc. The central peaks of the \ion{H}{$\alpha$} and \ion{H}{$\beta$} emission profiles exhibited subtle shifts towards the blue or red side, attributed to orbital motion and fluctuations in the accretion rate.  Using the double-peak features observed in the \ion{H}{$\alpha$} and \ion{H}{$\beta$} lines, we have estimated the accretion disc size to be \( R_{AD} = 3.10 \pm 0.04 \times 10^{12} \, \text{cm} \). The \textsc{cloudy} photoionization code is employed to model the quiescent phase spectra, allowing us to study the evolution of various physical parameters such as temperature, luminosity, hydrogen density, elemental abundances, accreted mass, and accretion rate. The central ionizing sources exhibit temperatures in the range of $1.05 - 1.80~\times 10^4$ K and luminosities between $0.10 - 7.94~\times 10^{30}$ \ergs. The mean accretion rate, calculated from the model, is $\sim$ $1.25 \times 10^{-8} M_{\odot}$ yr$^{-1}$. The model results reveal that the accretion rate rose substantially in the later phase. The accreted mass in the 16 months, preceding the 2021 outburst exceeds 47\% of the critical mass, and more than 88\% of the critical mass was accreted in the last three years.
\end{abstract}



\section{Introduction}
The Recurrent Nova (RN) RS Ophiuchi (RS Oph) is a symbiotic nova belonging to a rare class of binary star systems, consisting of a massive white dwarf ($M_{WD}=1.20-1.40~M_{\odot}$) and a red-giant (RG) of type M0--2 III with a mass range of $0.68-0.80~M_{\odot}$ \citep{2006Natur_Sokoloski, 2009A&A_Brandi, 2017ApJ_Mikolajewska,2007NewAR_Parthasarathy, 2007ApJ_Hachisu, 2011ApJ_Osborne, 2020MNRAS_Mondal, 2022MNRASPandey}.  The binary has an orbital period, $P_{\text{orb}}$, of 453.60 $\pm$ 0.40 days \citep{2009A&A_Brandi}.  The distance to RS Oph has been estimated by various scholars, with values ranging from 0.4 kpc to 5.8 kpc (see \citet[and references therein]{2008ASPC_401_Barry}). Some of the most widely cited estimates are: \(d = 1.6 \, \text{kpc}\) by \citet{1986ApJ_Hjellming}, \(d = 2.45 \pm 0.37 \, \text{kpc}\) by \citet{2008ApJ_688_559R_Rupen}, \(d = 3.1 \pm 0.5 \, \text{kpc}\) by \citet{2008ASPC_401_Barry}, and the GAIA DR-3 estimates: \(d = 2.68^{+0.17}_{-0.15} \, \text{kpc}\) \citep{2022A&A_666L_Munari}, \(d = 2.44^{+0.08}_{-0.16} \, \text{kpc}\) (geometric), and \(d = 2.44^{+0.22}_{-0.21} \, \text{kpc}\) (photogeometric) \citep{2023AJ_166_269B_Bailer-Jones}. In this system, the white dwarf (WD) accretes material from the RG's stellar wind, leading to a buildup of hydrogen on the WD's surface. When the accumulated hydrogen reaches a critical pressure, it ignites in a thermonuclear runaway (TNR), resulting in a nova outburst \citep{Gehrz1998PASP}. Previously RS Oph has undergone  nine repeated outbursts in 1898, 1907, 1933, 1945, 1958, 1967, 1985, 2006 \citep{2010ApJS_Schaefer}, and 2021 \citep{2022MNRASPandey}. However, the 1907 and 1945 outbursts lack full confirmation due to their alignment with the sun \citep{2004IAUC_Schaefer,2010ApJS_Schaefer}. The recurrence of these outbursts is interspersed with quiescent periods lasting approximately between 9 and  21 years \citep{2010ApJS_Schaefer}. 
 
The massive white dwarf, in conjunction with its high mass-transfer rate of  ($\sim~2.00 \times 10^{-7}M_{\odot}~\text{yr}^{-1}$) \citep{1977MNRAS_Walker, 2016MNRAS_Booth}, provides compelling evidence that RS Oph is a likely candidate for a Type Ia supernova. The gradual increment in the WD's mass, due to the accumulation of about 10\% of the accreted matter during each quiescent stage \citep{2000ApJ_Hachisu}, supports this likelihood. The recent study by \citet{2024ApJ_Starrfield} also showed a gradual increase in the WD's mass after outbursts, as the ejected mass appeared to be less than the accreted mass in each epoch of the hydrodynamical simulations conducted. Similarly, \citet{2008NewAR..52..386H_Hernanz} conducted an evolutionary model indicating that the ejected mass is smaller than the accreted mass. Consequently the WD mass may eventually exceed the Chandrasekhar limit ($M_{ch}=1.40~M_{\odot}$) sometime in the future \citep{2011ApJ_Osborne}. The net increasing rate of the WD mass has been calculated to be $1.20 \times 10^{-8}~M_{\odot}~\text{yr}^{-1}$ \citep{2000ApJ_Hachisu}, further supporting the premise that RS Oph holds the potential for evolving into a Type Ia supernova.
 
Several systematic multi-wavelength observations and studies of RS Oph have been conducted, mainly after the 1985 outburst. For example, $\gamma$-ray \citep{2022NatAs_6_689A_Acciari, 2022Sci_376_77H_HESS}, X-ray \citep{2006ApJBode, 2006Natur_Sokoloski, 2011ApJ_Nelson, 2022MNRAS.514.1557PPage}, ultraviolet \citep{1985ESASP_236_281C_Cassatella, 2011ApJ_Nelson}, optical \citep{2006CBET_403_1B_Buil, 1999A&A_Anupama, 2009NewA_14_336S_Skopal, 2021ATel14838_Taguchi}, infrared \citep{1988MNRAS_234_755E_Evans, 2006CBET_730_Das, 2008BASIP_25_50D_Das, 2006ApJ_653L.141D_Das, 2009MNRAS_Banerjee, 2007MNRAS.374L_1E_Evans, 2021ATel14866_Woodward}, and radio \citep{2006Natur_442_279O_OBrien, 2007ApJ_667L_171K_Kantharia, 2008ApJ_688_559R_Rupen}.

The comprehensive analysis of optical spectroscopy, particularly from 1958 onward \citep{1963AnAp_Eskioglu, 1967BAN_Pottasch}, revealed complex emission line profiles during outbursts, which are crucial for understanding the dynamics of the ejected material and its interaction with the surrounding environment. These profiles, including P Cygni profiles, indicate the presence of high-velocity winds and mass ejection from the system \citep{2006ApJBode, 1999A&A_Anupama}.  The UV observations in 1985, revealed significant differences in the emission line profiles of RS Oph compared to other classical novae \citep{1985ESASP_236_281C_Cassatella}. They reported the early appearance of high ionization species in the UV range. The complex UV emission line profiles are also crucial in  studying the dynamics of the ejection and the geometry of the system. \citet{1988MNRAS_234_755E_Evans, 2006ApJ_653L.141D_Das} made IR observation and reported the rare detection of an infrared shock wave as the nova ejecta plows into the preexisting wind of the secondary in the RS Oph system consisting of a WD primary and a RG secondary. \citet{1985Natur_315_306P_Padin}, reported the first detection of the radio emission from RS Oph, and suggested a non-thermal origin of high brightness temperature. RS Oph has been detected as a very strong soft X-ray source using the European X-ray Observatory Satellite (EXOSAT) \citep{1987rorn_Mason}. The early radio and X-ray emissions from RS Ophiuchi are predominantly non-thermal, resulting from the interactions between the shock waves and the surrounding nebular material \citep{2009MNRAS_Eyres, 2024MNRAS_Nayana}. \citet{MAGIC:2023wxo_Abe} reported the detection of very high-energy (VHE) gamma rays at a significant level of 13.2 during the first four days of RS Oph using the MAGIC telescopes. In August 2021, the Fermi Large Area Telescope, H.E.S.S., and MAGIC all detected GeV and TeV $\gamma$-ray emission from the 2021 outburst of the RN RS Ophiuchi \citep{2023ApJ_947_70D_Diesing}. This marks the first observation of very high-energy $\gamma$-rays from a nova, opening a new avenue for studying particle acceleration. Both H.E.S.S. and MAGIC attributed the observed $\gamma$-rays to a single, external shock.

Relative to the outburst phase of RS Oph, its quiescent phase has been studied to a lesser extent. Some of the studies conducted on it are; \citep{2009A&A_Brandi, 1999A&A_Anupama, 2018MNRAS_Zemko, 2020MNRAS_Mondal}. The quiescent phase of a nova is crucial for understanding various characteristics of nova eruptions and their development. This includes exploring interactions between binary components, the impact of eruptions on accretion discs and mass-transfer rates, recurrence intervals, and inter-class correlations among cataclysmic events \citep{2008BASIPKamath, 2020MNRAS_Mondal}.  This study aims to investigate the temporal evolution of the chemical and physical characteristics of the RS Oph system during its quiescent phase from 2008 to 2021.

We describe our datasets in Section \ref{DS}. In Section \ref{osa}, we discuss the spectral characteristics observed during quiescence between the 2006 and 2021 outbursts. Sections \ref{pma_rsoph} and \ref{MR} detail the model analysis and technique, respectively, using the \textsc{cloudy} photoionization tool. In Section \ref{RDNew}, we present the results and discussion of the model, and finally, we provide our conclusions in Section \ref{C}.

\section{Data Set} \label{DS}
We selected eight spectra spanning a time range of $\sim$ 13 years, from 2008 Feb. 22.38 UT to 2021 Mar. 8.15 UT, with an interval of about two years. These spectra cover the wavelength range of $\sim$ 3900 to 7500 \AA~ and have a resolution range of $\sim$ 620 to 14,000. They were selected for spectral analysis purposes, encompassing the quiescent period between the two consecutive outbursts (2006 and 2021). All the spectra are normalized to the line flux of \ion{H}{$\beta$}, and corrected for reddening by E(B-V)= 0.73 \citep{1987Ap&SS_Snijders, 2022MNRASPandey}.

 For the present study, we used spectroscopic data  available in  Astronomical Ring for Access to Spectroscopy Database (ARAS Database)\footnote{ \url{https://aras-database.github.io/database/novae.html}} \citep{2019Teyssier}, Stony Brook/SMARTS Atlas of (mostly) Southern Novae \footnote{\url{http://www.astro.sunysb.edu/fwalter/SMARTS/NovaAtlas/rsoph/rsoph.html}}\citep{2012PASPWalter}, and Astrosurf Recurrent Nova\footnote{\url{http://astrosurf.com/buil/us/rsoph/rsoph.htm}}. The first two spectra, from 2008 Feb. 22.384 to 2010 Aug. 15.824, were obtained from SMARTS and Astrosurf and were observed by Stony Brook (SB) and Christian Buil (BUI) using the Cerro Tololo Inter-American Observatory (CTIO) and the Castanet-Tolosan Observatory (CTO), respectively. The remaining six of the eight selected spectra were obtained from the ARAS database and correspond to the following observation dates: 2012 Jun. 26.937, observed by Christian Buil (BUI) using the Castanet-Tolosan Observatory (CTO); 2014 Jul. 24.947, observed by David Boyd (DBO) using the West Challow Observatory (WCO); 2016 Aug. 20.929, observed by Joan Guarro Flo (JGF) using the Santa Maria de Montmagastrell Observatory (SMMO); 2018 Jul. 20.021, observed by Tim Lester (LES) using the Mill Ridge Observatory (MRO); 2020 Apr. 06.339, observed by Tim Lester (LES) using the Mill Ridge Observatory (MRO); and 2021 Mar. 08.150, observed by Pavol A. Dubovsky (PAD) using the Vihorlat National Telescope (VNT). The Stony Brook/SMARTS Atlas database contains both spectroscopic and photometric data obtained since 2003. Astrosurf is an online astronomical platform where individuals share resources and observational data. The ARAS Symbiotics Project is composed of a cluster of compact telescopes, with diameters spanning from 20 cm to 60 cm. These telescopes are outfitted with spectrographs featuring resolutions ranging from R $\sim$ 500 to 15,000. The instruments cover a wavelength spectrum from 3600 {\AA} to $\sim$ 9000 {\AA} and are specifically designed for monitoring eruptive variable stars.  This database facilitate systematic studies of the nova phenomenon and correlative studies with other comprehensive data sets. We also utilized a spectrum of an M2III type star, obtained from the European Southern Observatory (ESO) website\footnote{\url{https://www.eso.org/sci/facilities/paranal/decommissioned/isaac/tools/lib.html}}, to match the absorption features in the spectra originating from the secondary star. The instrumentation and observation details for each observatory and the log of observations are presented in Table \ref{tab:T1_RSoph}.
 
\begin{table*}
	\caption{Log of selected optical spectral observation of RS Oph, during its quiescent stages between the 2006 and 2021 outbursts.   
	}
	\label{tab:T1_RSoph}
	\centering
	\setlength{\tabcolsep}{5pt}
	\begin{tabular}{l c c c c c c c c c r}
		\toprule
		Date (UT) & $t_0^a$ (days)&$t_q^b$ (days)&Source&Observer &Observatory& Spectrograph& Camera & R$^{c}$ & Coverage (\AA)&TTE$^{d}$(s)\\ 
		\midrule 
		2008 Feb. 22.384&739.550&300.00&SMARTS&SB$^e$&CTIO$^1$&venerable RC&1K CCD&-&2720-9558&300\\
		2010 Aug. 15.824&1644.99&1205.00&Astrosurf&BUI$^f$&CTO$^2$&LISA&QSI583&620&3829-7317&4200\\
		2012 Jun. 26.937&2326.11&1886.00&ARAS&BUI$^f$&CTO$^2$&LISA&Atik314L$^+$&1000&3829-7317&2113\\
		2014 Jul. 24.947&3084.12&2644.00&ARAS&DBO$^g$&WCO$^3$&LISA&SXVR-H694&870&3800-7591&4441\\
		2016 Aug. 20.929&3842.12&3402.00&ARAS&JGF$^h$&SMMO$^4$&LHIRES&ATIK 460EX&1101&3917-7475&4019\\
		2018 Jul. 20.021&4540.09&4100.00&ARAS&LES$^i$&MRO$^5$&echelle&ASI1600mm&12000&4031-7950&8590\\
		2020 Apr. 06.339&5166.50&4726.00&ARAS&LES$^i$&MRO$^5$&echelle&ASI1600mm&14000&4031-7955&7397\\
		2021 Mar. 08.150&5502.32&5062.00&ARAS&PAD$^j$&VNT$^6$&LISA&Atik 460ex&824&4000-7500&1822\\
		\bottomrule
	\end{tabular}\\
	{\raggedright Note: $^{(a)}$Number of days counted from outburst date ($t_0$) (2006 Feb. 12.83 UT), $^{(b)}$Number of days counted from the start of the possible date of the quiescent phase ($t_q$) (i.e., 2007 April 26), $^{(c)}$Resolution, $^{(d)}$Total Time of Exposure, $^{(e)}$Stony Brook using $^{(1)}$Cerro Tololo Inter-American Observatory, $^{(f)}$Christian Buil using $^{(2)}$Castanet-Tolosan Observatory, $^{(g)}$David Boyd using $^{(3)}$  West Challow Observatory, $^{h}$Joan Guarro Flo using $^{(4)}$Santa Maria de Montmagastrell Observatory, $^{(i)}$Tim Lester using $^{(5)}$Mill Ridge Observatory, and $^{(j)}$Pavol A. Dubovsky using $^{(6)}$Vihorlat National Telescope. 
		\par}
\end{table*}

\section{General descriptions of the spectra} \label{osa}
\begin{figure*}
	\centering
	\includegraphics[scale=0.82]{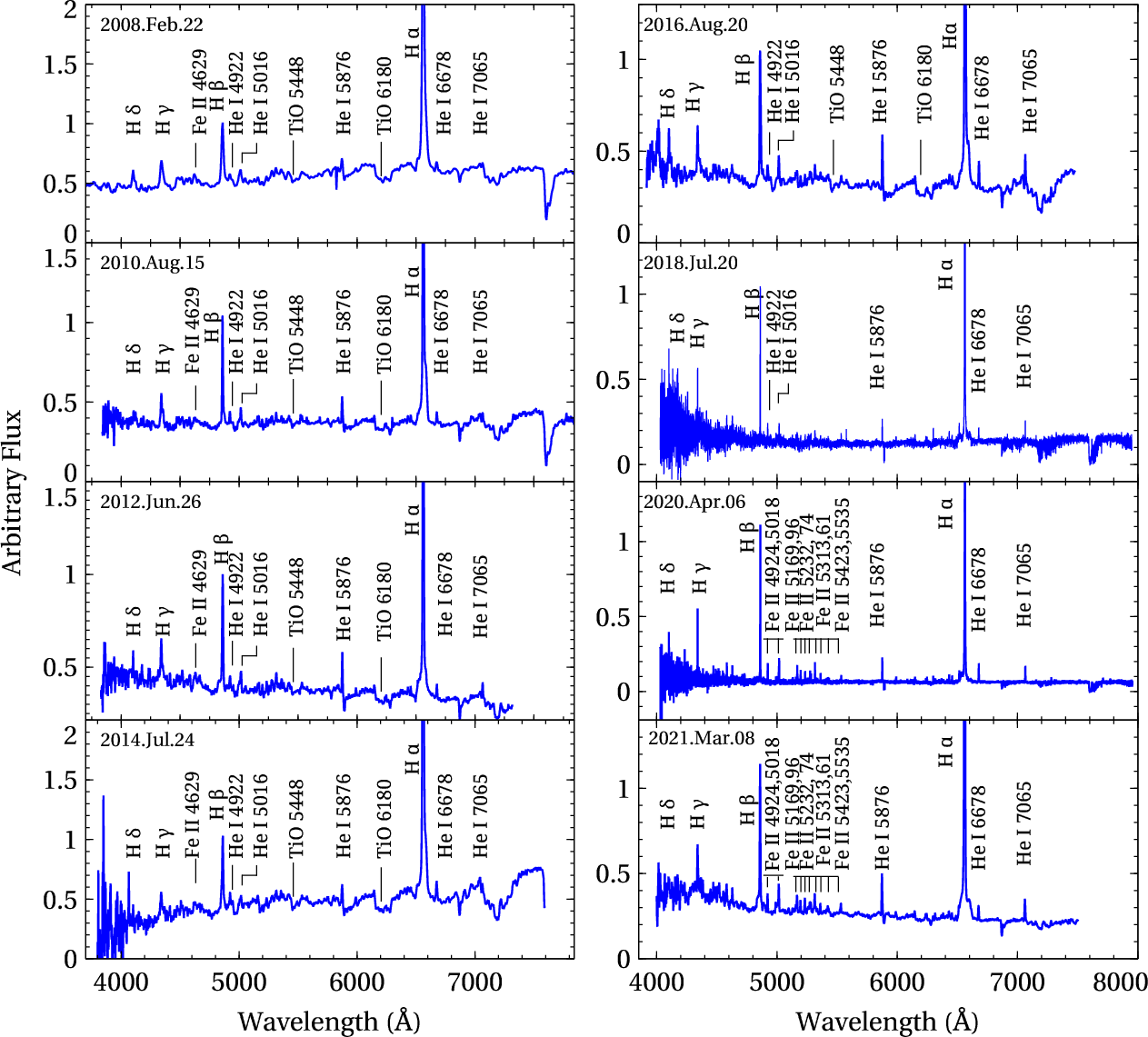}
	\caption{ Dereddened spectra of RS Oph, illustrating the spectroscopic evolution from days 739 to 5504 following the 2006 outburst. All the prominent lines are labeled.  The ordinate represents arbitrary normalized flux to \ion{H}{$\beta$} line whereas the abscissa represents the wavelength in \AA~ unit.   All spectra are corrected for reddening by a factor of E(B-V)=0.73.  
	}
	\label{fig:spec_evolution}
\end{figure*}

Fig. \ref{fig:spec_evolution} illustrates the evolution of the quiescent spectra of RS Oph from day 739.55 (2008 Feb. 22.38 UT) to day 5502.32 (2021 Mar. 8.15 UT), after the 2006 outburst ( on 2006 Feb. 12.83 UT), covering $\sim$ 4762 days of the quiescent period. The respective observation dates are indicated at the top-left corner of each panel. For this study, we selected eight spectra with nearly two-year intervals, except for the last one, which has only a one-year difference.  These spectra show strong and broad emission lines attributed to hydrogen, helium, and iron (see Table \ref{tab:fwhm}). The  spectra are mostly dominated by Balmer lines (from \ion{H}{$\alpha$} to \ion{H}{8}) and He I (4922, 5016, 5876, 6678, and 7065 \AA) lines. The presence of iron also becomes clear, especially in the later phases of the quiescent stage. \ion{Fe}{ii} (4233, 4491, 4584, 4924, 5018 \AA) are some of the iron lines that appear in the spectra. The absence of higher ionization lines in the spectrum is likely due to the high-energy photons from the accretion disc being absorbed by the surrounding material. These photons are then re-emitted at lower energies, resulting in the softening of the radiation. The emission lines observed in these spectra during quiescent phase appear broad due to the expanded discs \citep{1989JApAAnupama}. Intense optical \ion{Fe}{ii} emission lines originate from the outer, lower density portion of the disc (see Sec. \ref{ec}). The strong \ion{TiO}{} absorption features at 5448 and 6180 \AA~ originate from the cool secondary star, indicating the secondary star is a cooler star. The appearance of optical flickering is considered a sign of the resumption of the accretion process \citep{1998A&A...338..988Z_Zamanov}. \citet{2007MNRAS_Worters, 2008ASPC_Worters} reported that the accretion process resumed on day 241, based on their photometric observations, which showed strong optical flickering \citep{2001MNRAS.326..553S_Sokoloski}. On the other hand, \citet{2018MNRASMondal, 2020MNRAS_Mondal} reported that the accretion process resumed approximately on day 250 following the 2006 outburst. However, the full quiescent phase began on April 26, 2007, which corresponds to day 440 following the 2006 outburst \citep{2018MNRASMondal}. In Table \ref{tab:T1_RSoph}, we present the number of days from the onset of the quiescent phase ($t_q$) until the date the spectra in Fig. \ref{fig:spec_evolution} were taken.
\defcitealias{smith2014}{Paper~I}
 
\subsection{Emission Line profiles}\label{ELP}
\begin{figure}
	\centering
	\includegraphics[scale=0.65]{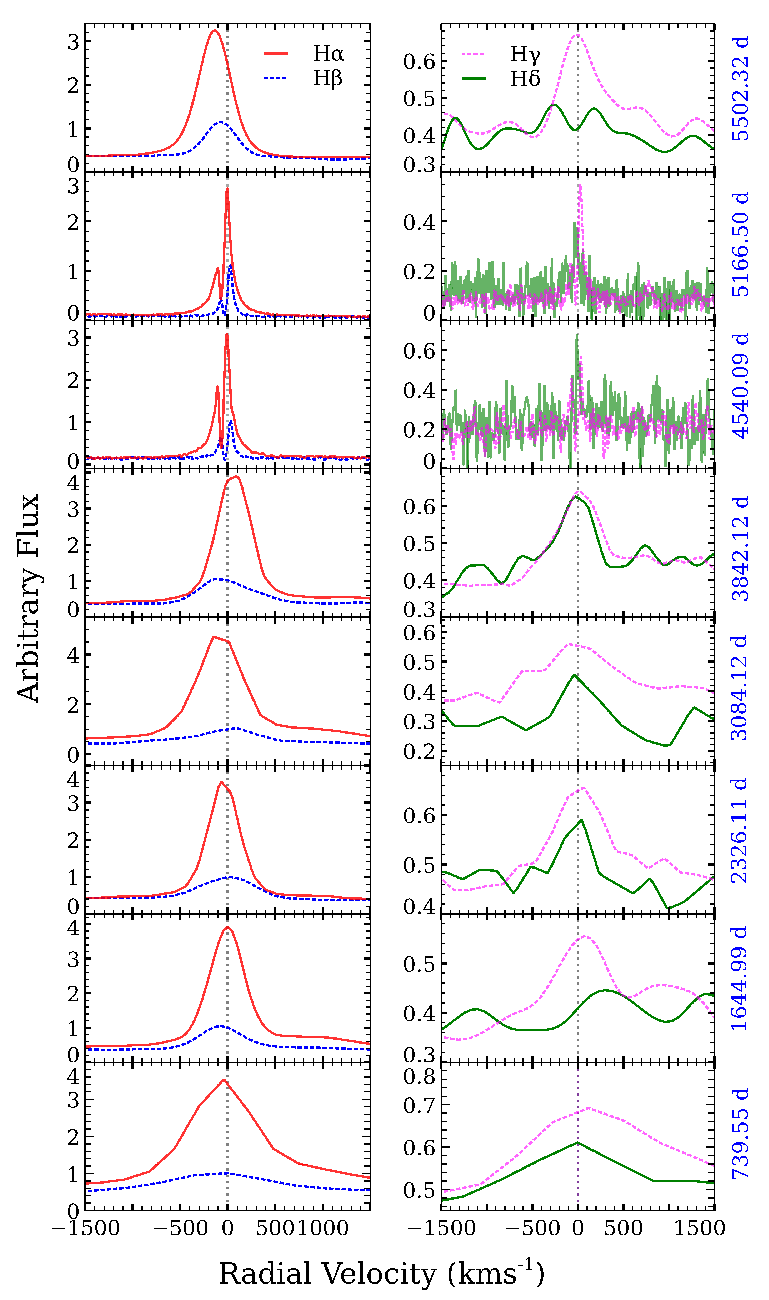}
	\caption{  Plot of \ion{H}{$\alpha$}, \ion{H}{$\beta$}, \ion{H}{$\gamma$}, and \ion{H}{$\delta$} emission  line profiles of RS Oph during the quiescence phase between the 2006 and 2021 eruption. The legends on the top panels of each column apply to all the panels below, and the number of days provided on the right side of the second column serve as labels for the corresponding rows in the left column as well, representing number of days counted from outburst date ($t_0$) (2006 Feb. 12.83 UT).   
	}
	\label{fig:21x}
\end{figure}

The emission line profiles of the four strongest Balmer lines (\ion{H}{$\alpha$}, \ion{H}{$\beta$}, \ion{H}{$\gamma$}, and \ion{H}{$\delta$}) are depicted in the two panels of Fig. \ref{fig:21x}. The left column shows the profiles of the \ion{H}{$\alpha$} and \ion{H}{$\beta$} lines, while the right column shows the profiles of the \ion{H}{$\gamma$} and \ion{H}{$\delta$} lines. Both columns encompass the profiles from eight distinct epochs. The \ion{H}{$\alpha$} and \ion{H}{$\beta$} profiles of 2018 Jul. 20.02 UT (day 4540) and 2020 Apr. 6.34 UT (day 5166.50), acquired through a high-resolution telescope, distinctly exhibit a central deep absorption feature which cut the broad emission line into two adjacent peaks (see the left column of Fig. \ref{fig:21x}). This is probably due to a slow, very dense wind in the system \citep{1993A&AS_Van_Winckel, 1999A&A_Anupama}. These double-peak features of \ion{H}{$\alpha$} and \ion{H}{$\beta$} suggest that they originate from the accretion disc \citep{1986MNRAS_218_761H_Horne, 2024arXiv240511506ZZamanov}. This feature, previously observed in various prior outbursts \citep{2009A&A_Brandi, 2011BlgAJ_Zamanov, 2014MNRAS_Worters, 1993A&AS_Van_Winckel, 1999A&A_Anupama}. The red-side peak of both \ion{H}{$\alpha$} and \ion{H}{$\beta$} emission lines of both epochs, appeared stronger than the blue one in both epochs, consistent with observations during the quiescent period of the 1985 outburst of RS Oph \citep{2009A&A_Brandi, 1993A&AS_Van_Winckel}.

The core of the \ion{H}{$\alpha$} line displays subtle positional shifts toward either the blue or red side (see Fig. \ref{fig:21x}). For instance, on days 739.55, 2326.11, 3084.12, 4540.09, and 5502.32, the core has been shifted blueward by -26, -33, -67, -15, and -131 \kms, respectively, while on days 1664.99 and 3842.32, the core has been shifted to the red by +5 and +93 \kms, respectively. The central deep absorption noticed on these two days (4540.09 and 5166.50) have been also shifted bluewards by  -69 and -73 \kms. The reason for these shifts of peaks of emission lines could be the orbital motion of the primary and secondary stars around their common center of mass. This orbital motion induces periodic Doppler shifts in the emission lines \citep{1986MNRAS_218_761H_Horne}. Another possible reason for the observed positional shift of the peaks could be fluctuations in the accretion rate. \citet{1999A&A_Anupama} reported that accretion rate fluctuations were the primary reason for various variabilities, including changes in emission flux. A similar phenomenon was observed in nova V3890 Sgr \citep{2018MNRAS_Zemko}. Comparable shifts in the central position have also been noted in other Balmer lines, such as \ion{H}{$\beta$}, \ion{H}{$\gamma$}, and \ion{H}{$\delta$}. Unfortunately, we didn’t find them shifting in fully similar sequences with \ion{H}{$\alpha$}, especially for \ion{H}{$\gamma$} and \ion{H}{$\delta$}. This discrepancy could possibly be due to the higher-ionized Balmer lines being influenced by de-blending with nearby \ion{Fe}{} lines or a lower signal-to-noise ratio. 

The most noticeable characteristic observed in the emission lines of the selected spectra in this study is a variations in the width of the Balmer lines over time. Table \ref{tab:fwhm} presents the FWHM values of \ion{H}{$\alpha$} and \ion{H}{$\beta$} for the eight spectra. The table clearly shows that the emission line profiles were widest on 22 Feb. 2008 compared to any other spectra captured later. Despite a slight increase on 24 Jul. 2014, the FWHM values of both emission profiles decrease monotonically until 06 Apr. 2020. Contrary to the general decrease in line width over time, significant broadening was observed on 08 Mar. 2021, about five months before the next outburst. This indicates that the line profiles were strongly influenced by the resolving power of the telescopes used. This narrowing of line widths could possibly be attributed to the slowing down of the remaining shell ejecta from the previous outburst, likely caused by interactions with the surrounding interstellar medium and a consequent decrease in Doppler broadening. A similar deceleration of shell ejecta has been observed in T Pyx \citep{2010ApJ...708..381S_Schaefer}. However, due to the varying spectral resolution of the instruments used, which directly affects the line width, we are unable to make a conclusive statement regarding the cause of the observed variations in the widths of these Balmer lines. The broadening observed in the last epoch could potentially be due to an increased accretion rate as the system approaches the critical limit, resulting in higher velocity dispersion of the accreted material in the accretion disc. Additionally, the increased mass accreted onto the white dwarf may generate turbulence in the accretion disc, further contributing to the broadening \citep{warner1995, 2002apa_book_Frank}. Our photoionization modeling in Sec. \ref{amr_rsoph} confirmed a significant increase in the accretion rate.

\begin{table}
	\centering
	\caption{ FWHM values of most prominent emission lines (\ion{H}{$\alpha$} and \ion{H}{$\beta$}) in all quiescent phase spectra.}
	\label{tab:fwhm}
	\begin{tabular}{lccr}
		\hline
		Epochs&$\bigtriangleup t$&FWHM(\ion{H}{$\alpha$})&FWHM(\ion{H}{$\beta$})\\
		&(days)&(\kms)&(\kms)\\
		\midrule
		22 Feb. 2008&739.550&1114&1500\\
		15 Aug. 2010&1644.99&913 &633 \\
		26 Jun. 2012&2326.11&442 &724 \\
		24 Jul. 2014&3084.12&658 &1021 \\
		20 Aug. 2016&3842.12&470 &743 \\
		20 Jul. 2018&4540.09&247 &127 \\
		06 Apr. 2020&5166.50&185 &130  \\
		08 Mar. 2021&5502.32&460 &449 \\
		\hline
	\end{tabular}\\
	{\raggedright  Note: $\bigtriangleup t$ is the time interval counted from outburst date ($t_0$) (2006 Feb. 12.83 UT) to each dates. Each FWHM are subjected to $\pm$ 10 to 30 \kms error, which is estimated by taking three measurements for each and averaging them.      \par}
\end{table}

\subsubsection*{Disc size estimation from the \ion{H}{$\alpha$} and \ion{H}{$\beta$} double peaks} \label{DSize}
In Section \ref{ELP}, we discussed that the \ion{H}{$\alpha$} and \ion{H}{$\beta$} emission profiles showed double peaks on 2018 Jul 20.02 UT (day 4540) and 2020 Apr 6.34 UT (day 5166.50) (see Fig. \ref{fig:21x}). These features are attributed to emission lines originating from the disc. We assume the accretion disk surrounding the WD of RS Oph is a Keplerian disk, where the disk orbits the WD under the influence of gravity, similar to TCrB \citep{2024arXiv240511506ZZamanov}. By using the separation of double peaks \(\bigtriangleup V_{\ion{H}{$\alpha$}}\) on the lines originated from the Keplerian disc, the outer radius \(R_{AD}\) can be estimated using the relation given by \citep{1972ApJ_171_549_Huang}:
\begin{equation}\label{eq1}
	\bigtriangleup V_{\ion{H}{$\alpha$}} = 2\sin i\sqrt{\dfrac{G M_{WD}}{R_{AD}}}
\end{equation}
where \( G \) is the gravitational constant (\(6.67 \times 10^{-11} \, \text{m}^3 \text{kg}^{-1} \text{s}^{-2}\)), \( M_{WD} \) is the mass of the WD in RS Oph, and \( i \) is the inclination angle of the disc axis to the line of sight. We adopt the value \( M_{WD} = 1.35 M_{\odot} \) \citep{2007ApJ_Hachisu}.
	The inclination angle of RS Oph has been estimated by various researchers as follows: \( i = 30^{\circ} \) \citep{2022MNRASPandey}, \( i = 39^{+1}_{-9} \)$^{\circ}$ \citep{2012MmSAI_83_762_Ribeiro}, \( i = 50 \pm 1^{\circ} \) \citep{2009A&A_Brandi}, and \( i = 30 - 40^{\circ} \) \citep{1994AJ1082259_Dobrzycka}. Among these estimates, we selected the two extreme values of the inclination angle, \( i = 30^{\circ} \) and \( i = 51^{\circ} \), to determine the most probable range of the disc radius.

	The peak separations for \ion{H}{$\alpha$} and \ion{H}{$\beta$} were found to be \( 98 \pm 0.28 \, \text{km/s} \) and \( 101 \pm 0.83 \, \text{km/s} \), respectively. Applying these values in Eq. \ref{eq1}, we found \( R_{AD1} = 1.87 \pm 0.01 \times 10^{12} \, \text{cm} \) for \ion{H}{$\alpha$} and \( R_{AD2} = 1.76 \pm 0.03 \times 10^{12} \, \text{cm} \) for \ion{H}{$\beta$}, for \( i = 30^{\circ} \). Similarly, for \( i = 51^{\circ} \), we found \( R_{AD1} = 4.51 \pm 0.03 \times 10^{12} \, \text{cm} \) for \ion{H}{$\alpha$} and \( R_{AD2} = 4.25 \pm 0.07 \times 10^{12} \, \text{cm} \) for \ion{H}{$\beta$}. Taking the average of each result gives the radius of the accretion disc as \( R_{AD} = 3.10 \pm 0.04 \times 10^{12} \, \text{cm} \).

\section{Photoionization Model Analysis} \label{pma_rsoph}
We used the 2023 released version of the \textsc{cloudy} code (\textsc{C23})\footnote{\url{https://trac.nublado.org/}}; \citep{2023RMxAA_Chatzikos} to model the quiescent spectra of RS Oph. \textsc{cloudy} has been effectively applied to study novae including  the quiescent stage of a nova \citep{2015NewA_Das, 2018MNRASMondal, 2019A&A_Pavana, 2020MNRAS_Mondal}. \textsc{cloudy} simulates the physical conditions of non-equilibrium gas clouds exposed to an external radiation field by using detailed microphysics. It predicts the emission-line spectrum based on assumptions about the gas's physical conditions (ionisation, density, temperature, and chemical composition). The photoionization code \textsc{cloudy } uses a set of input parameters to compute the ionization, thermal and chemical state of a non-equilibrium gas cloud, illuminated by a central source, and it predicts the resulting spectra. The input parameters include the temperature (T) in Kelvins and luminosity (L) in \ergs~of the central ionising source, hydrogen number density ($n$) in cm$^{-3}$, filling factor, covering factor, elemental abundances, and inner and outer radii (cm) of the surrounding accretion disc. To generate synthetic model spectra, we incorporated all of these input parameters in our model along with the abundances of only those elements whose emission lines are observed in the spectra, while other elements are kept at their solar values \citep{2010ApGrevesse}. In \textsc{cloudy}  hydrogen density and filling factor varies radially as $	n(r)= n(r_o)(r/r_o)^{\alpha}~\text{cm}^{-3}$ and $	f(r) = f(r_o)(r/r_o)^{\beta}$, respectively, where $\alpha$ and $\beta$ are the exponents of the power laws, and $r_0$ is the inner radius. The density of the disc is controlled by a hydrogen density parameter with a power-law density profile and an exponent of -2 because it provides a steady mass per unit volume throughout the model disc ($\dot{\textrm{M}}=$ const). The ratio of the filled to vacuum volumes in the disc are set to 0.1, which is the value used in other recent studies using \textsc{cloudy}  \citep{2015NewA_Das, 2018MNRASMondal, 2020MNRAS_Mondal}.

The goodness of our model fit is estimated from the  $\chi^2$ and $\chi_{red}^2$ (reduced-$\chi^2$) of the
model; using $\chi^2=\sum_{i=1}^{n}\frac{(M_i-O_i)^2}{\sigma^2}$, and  $ \chi_{red}^2=\frac{\chi^2}{\nu}$ respectively, where $O_i$ and  $M_i$ are the ratios of observed and modelled line fluxes to the \ion{H}{$\beta$} line flux, respectively, $\sigma_i$ is the error in the observed flux ratio, $\nu$ is the degrees of freedom given by $n-n_p$, n is the number of observed lines, and $n_p$ is the number of free parameters. An ideal model has a $\chi^2 \approx \nu$ \citep{2001MNRAS-Schwarz}.  Thus, the $\chi^2_{red}$ value of a model has to be low (typically in the range of 1-2) in order to be considered acceptable and well fitted. Normally, $\sigma$ ranges from 10 to 30 percent, depending on how strong it is relative to the continuum and whether it can blend with other lines in the spectrum \citep{Helton10}. 

\subsection{Modeling Procedure }\label{MR}
We modeled the first seven spectra (2008-2020), offering broader spectral coverage ($\sim$ 3900 to $\sim$ 7500 \AA~) and featuring a greater number of emission lines ($\sim$ 12 to 18 lines).We chose seven spectra to cover the entire quiescent period, with approximately two-year intervals. The chosen seven epochs are Epoch 1 (2008-Feb-22), Epoch 2 (2010-Aug-15), Epoch 3 (2012-Jun-26), Epoch 4 (2014-Jul-24), Epoch 5 (2016-Aug-20), Epoch 6 (2018-Jul-20), and Epoch 7 (2020-Apr-2020); corresponding to days 739, 1644, 2326, 3084, 3842, 4540, and 5166, respectively after the 2006 outburst.

\begin{figure}
	\centering
	\includegraphics[scale=0.33]{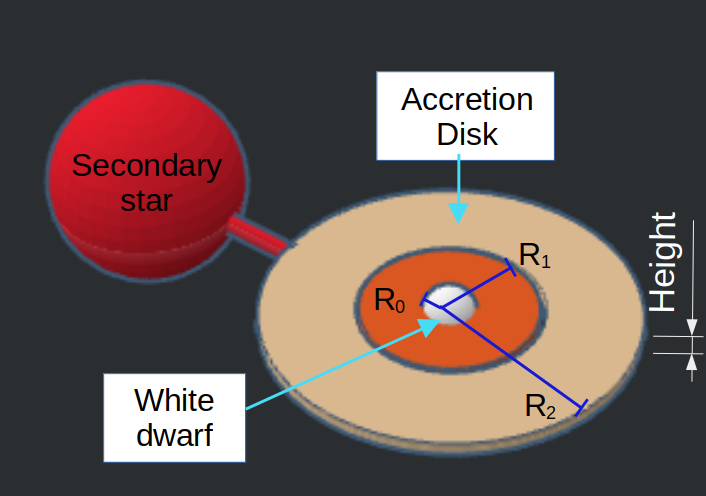}
	\caption{Schematic diagram of a nova system during the later phase of the quiescence stage (2018–2021). The accretion disc is assumed to be cylindrical in shape and has two components of density and temperature. The symbols $R_0$, $R_1$, and $R_2$ represent the radius of the WD (inner radius of the inner disc component), the outer radius of the inner disc component (the inner radius of the outer disc component), and the outer radius of the outer disc component, respectively.  }
	\label{fig:system}
\end{figure}

Initially, during the early years of the quiescent phase (from 2008-2016), we employed a one-component density profile across the radial distance of the accretion disc. However, as the nova  system approached to the next outburst (from 2018-2021), the accretion disc increased in size significantly, leading to density gradient which is responsible for the emergence of diverse spectral lines. To account for this, we divided the accretion disc into two components (see Fig. \ref{fig:system}) with distinct density and temperature profiles and adjusted them to match the emission features by multiplying with the corresponding covering factors. However, before introducing the two-component disc model, we ran several one-component models for the last two epochs as well. In these cases, we consistently encountered the challenge of underestimating various iron lines, a limitation that was clearly mitigated after employing the two-component model. In the two-component model, while slightly varying the temperature and keeping the luminosity constant in each component, we noticed that the fitting procedure was not very sensitive to small changes in luminosity. This could be attributed to the minimal variation in geometrical extension with the luminosity parameter  \citep{1976ApJStarrfield, 1978ARA&AGallagher, 2022MNRASPandey}. Iron lines observed in the spectrum were mostly generated from the outer, lower-density region of the disc, whereas the helium lines primarily originated from the inner, higher-density and higher-temperature region.   

In addition to the accretion disc, we also incorporated the primary (WD) and secondary stars into our model. The primary star's contribution was added to the synthetic spectrum by generating a blackbody for each corresponding temperature, which fit the continuum. The secondary star's contribution was included by using the spectrum of an M2 III type RG star. Inclusion of the secondary star's spectrum in our model fits the absorption features originating from the cool secondary star. The accretion disc was responsible for all the emission lines in the spectra, and we adjusted its parameters for the best fit. 

The inner radius \((R_{\text{in}})\) and outer radius \((R_{\text{out}})\) of the accretion disc  were calculated as follows.
 The \(R_{in}\) was calculated from the relationship between the radius and mass of the WD  (i.e., \(~M_{\text{WD}}^{(1/3)}~R_{\text{WD}}\) $\propto$ constant) \citep{Gehrz1998PASP}, and using the result that a WD of \(1 M_{\odot}\) has a radius of \(10^{8.9}\)cm \citep{1961ApJ_683H_Hamada}. The  \(~R_{\text{in}}~\) is equals to \(R_{WD}\), as the inner portion of the disc is in physical contact with the outer surface of the WD.  Therefore by equating the two relations we obtained \(R_{in}=10^{8.86}\) cm. The \(R_{out}\), was estimated using the relation \(R_{\text{AD}}(\text{max})/a=0.60/(1+q)\), where  \(R_{\text{AD}}(\text{max})\), \(a\) and \(q\) represent the maximum accretion disc radius (i.e., equivalent to \(R_{out}\)),  separation, and ratio between the primary and secondary stars, respectively \citep{Paczynski1977ApJ, 2008A&A_Lasota, 2023MNRAS_Sun}. We took the average of the lowest and highest possible masses of the secondary (0.68 - 0.8 $M_{\odot}$) \citep{2017ApJ_Mikolajewska}, which is \(0.74 M_{\odot}\). Using Kepler's third law and substituting the mentioned values along with the orbital period of 453.6 days \citep{2009A&A_Brandi}, we obtained the separation (\(a\)) to be \(2.22 \times 10^{13}\)cm, and the ratio (\(q\)) to be \(0.54\). By substituting these all in the equation above, we obtained \(R_{out}~=~10^{12.94}\) cm. This estimate of the disc size is consistent with our estimation of the disc size from the double peak features of \ion{H}{$\alpha$} and \ion{H}{$\beta$} in (Section \ref{DSize}). In our model of accretion disc we used \(R_{in}=10^{8.85}\)cm and \(R_{out}=10^{13}\)cm.  

The inner radii for all epochs, except for the outer components of the last two epochs where we have considered a two-component model, are the same. However, as the accretion disc continuously grows radially, the outer radii of each epoch had to vary. Therefore, in our model we made the outer radius a free parameter except for the second component of the last epoch. In the initial years of the quiescent stage (particularly from 2008-2016), the outer radii were varying almost linearly with time. However, in the last few years (2018-2020), the outer radii of each epoch were increasing more rapidly. Fig. \ref{fig:accdisc} illustrates this phenomenon effectively.

\begin{figure*}
	\centering
	\includegraphics[scale=0.82]{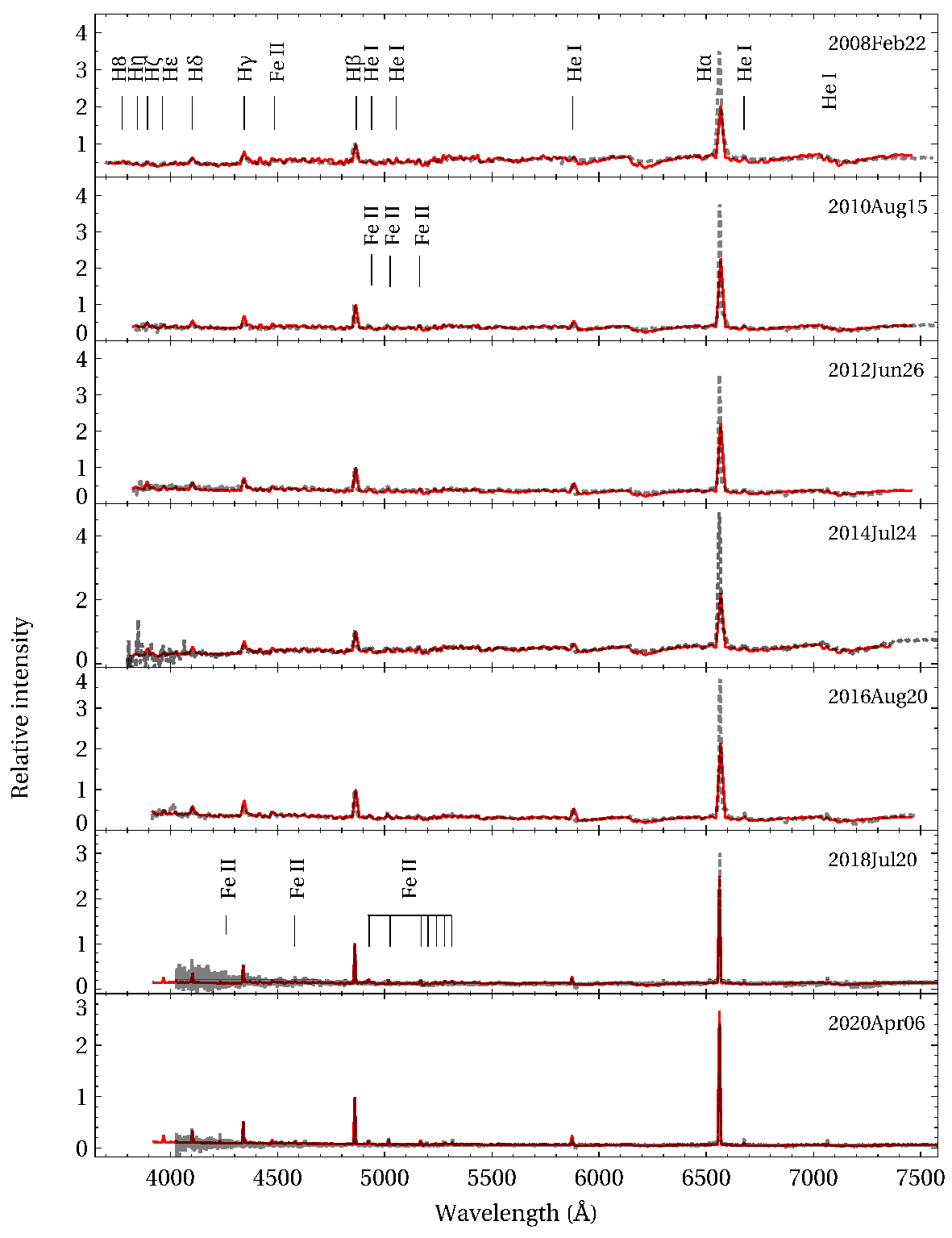}
	\caption{ Observed spectra during the quiescence phase of nova RS Oph (grey dashed line) alongside model-generated spectra (red solid line). Dates of observation for each epoch are indicated on the right side of each panel. Spectra have been normalized with respect to the \ion{H}{$\beta$} line. Prominent emission lines are highlighted in the figure. For more details, please refer to Section \ref{MR}.   
	}
	\label{fig:model}
\end{figure*}

From the best-fitting model, we have obtained values for various physical and chemical parameters of the RS Oph system during its quiescent phase. These parameters include the temperature and luminosity of the source, hydrogen number density, dimensions, and composition of the accretion disc formed on the surface of the WD (see Table \ref{tab:T2_results}). We have also estimated the uncertainties of the five free parameters: temperature, luminosity, hydrogen density, and the \ion{He}{} and \ion{Fe}{} abundances. From the measured uncertainties, we observed that the temperature varies within a narrow range, whereas the density exhibits a broader range of variation. This suggests that the disc properties are more sensitive to temperature variations than to any of the other free parameters. The uncertainties are presented in Table \ref{tab:T2_results}.  Eventually, we have calculated the chi-square values again In Fig. \ref{fig:model}, we present the best-fitting synthetic spectra overlaid on the observed spectra of RS Oph for seven epochs. Both the modeled and observed spectra are normalized to the line flux of \ion{H}{$\beta$}, and the observed spectra are dereddened by E(B-V)=0.73 \citep{1987Ap&SS_Snijders, 2022MNRASPandey}.
\defcitealias{smith2014}{Paper~I}

To minimize the number of free parameters, the density power, filling factor, and inner radius are held constant throughout, while only the outer radius of the second component in the last epoch (epoch 7) is held constant during the iterative process of fitting the observations. The hydrogen density, underlying luminosity, and effective blackbody temperature were varied. In addition, only the abundances of elements of observed lines were varied. All others are either set fully off or fixed at their solar values \citep{2010ApGrevesse}. We excluded certain elements from our model spectra—such as carbon, oxygen, nitrogen, neon, sodium, calcium, magnesium, and aluminum—due to their negligible presence in the observed spectra. Consequently, setting them to the solar abundance level, like many other elements, significantly affects the fitting. 

Following the procedure, we computed a set of synthetic spectra by simultaneously varying all the aforementioned input parameters in smaller increments across a broad sample space. The temperature ranged from $10^{3.5}$ to $10^5$ K, luminosity was varied between $10^{27}$ and $10^{35}$ \ergs, and the disc density spanned $10^{8}$ to $10^{13} \text{cm}^{-3}$, concurrently with the elemental abundances. Multiple test models were iterated across all epochs before arriving at the final model. Initial visual examinations were conducted, and model spectra that did not align with the observed spectrum were discarded. To assess the fit quality, we calculated the $\chi^2$ and $\chi^2_{\text{red}}$ values of the model, as discussed in section \ref{pma_rsoph}. A comparison of the relative fluxes for the best-fitting model-predicted lines and the observed lines during the early phase is presented in Table \ref{tab:T3_chi_square}, along with the corresponding $\chi^2$. We selected emission lines that appear in both observational and modeled spectra, for the calculation of $\chi^2$. To determine the line fluxes in individual emission lines, interactive flux measurements were performed by fitting Gaussians using the \textit{splot} task of the \textit{onedspec} package in \textsc{IRAF}.

\subsection{Results and Discussion}\label{RDNew}
\subsubsection{Temperature \& Luminosity} \label{TLD}
From the first epoch (2008) to the fifth (2016), we applied a one-component model. The temperature and luminosity of the system increased from $~1.05~ \times 10^{4}$ K and $~1.00~\times 10^{29}$ \ergs to $~1.20~ \times 10^{4}$ K and $1\times 10^{30}$ \ergs, respectively.  In the initial couple of years, the system was relatively cool and less luminous system, possibly indicating that it had fully entered the quiescent phase with minimal matter accretion at that moment. Over time, both the temperature and luminosity increased.  We opted not to vary the luminosity between the components considered in the last two epochs  as we observed that the fitting process was somewhat insensitive to small variations in luminosity (see Sec. \ref{MR}). Similar to the patterns observed in the previous epochs, the temperature and luminosity continued to increase in these cases as well (see the values in columns 7-10 of Table \ref{tab:T2_results}). 

\subsubsection{Density \& Radius} \label{DR}
Our model shows that the hydrogen density in the accretion disk increased from $3.16 \times 10^{9}$ \text{cm}$^{-3}$ to $6.31 \times 10^{10}$ \text{cm}$^{-3}$ from 2008 to 2016. After 2016, we used two density components: the inner density increased from $6.31 \times 10^{10}$ \text{cm}$^{-3}$ to $3.16 \times 10^{11}$ \text{cm}$^{-3}$, while the outer density decreased from $6.31 \times 10^{10}$ \text{cm}$^{-3}$ to $3.16 \times 10^{10}$ \text{cm}$^{-3}$. This clearly demonstrates that as the accretion disk forms, the inner region becomes denser than the outer region due to higher gravitational compression in the lower layers of the disk. In addition to this the temperature  and viscous heating get lesser in the outer portion of the disc, which has a direct impact on the density distribution \citep{2021A&A...646A.102L_Labdon, 2024MNRAS.527.9655A_Alarcon}.

The model also show that the outer radius of the accretion disk increased from $3.20 \times 10^{9}$ cm to $6.31 \times 10^{11}$ cm during the period from 2008 to 2016. From 2016 to 2020, it further increased from $6.31 \times 10^{11}$ cm to $1.00 \times 10^{13}$ cm. This indicates that the radial expansion of the accretion disk was faster in the later stages of the quiescent phase compared to the earlier stages. The study conducted by \citet{2024arXiv240511506ZZamanov} showed how the disc size varied over a duration of six months. Although the results indicated a pattern of oscillation between larger and smaller sizes, the trend suggests that the disc size is increasing over time. 

\subsubsection{Elemental composition} \label{ec}
The spectra show prominent lines of hydrogen, helium, and iron. All spectra are primarily dominated by Balmer lines, such as \ion{H}{$\alpha$}, \ion{H}{$\beta$}, \ion{H}{$\gamma$}, and \ion{H}{$\delta$}. These lines are the strongest in all spectra, indicating that hydrogen is the major constituent of the disc.

Our model reveal a significant decrease in the abundance of helium throughout the quiescent stage of the nova RS Oph. For example, the \ion{He}{}/\ion{He}{$\odot$} ratio was 2.4 in the first epoch but decreased to the solar value on epochs 6 and 7. The possible reason for the decrease in \ion{He}{} abundance over time during the quiescent phase could be the dilution effect and material mixing. During this phase, hydrogen-rich material from the secondary star is steadily accreted onto the white dwarf. This continuous supply dilutes the proportion of helium in the disc, leading to an apparent decrease in its abundance. \citet{2021ApJ9145_Sparks} suggested that the accreted material undergoes dilution and mixing, including convection and thermohaline mixing, which diminishes abundance enhancement of elements over time. Similarly, \citet{Iben_Fujimoto_2008} reported that the critical amount of accreted helium increases as the accretion rate decreases, which is fully consistent with our results. Additionally, \citet{2018MNRASMondal} found that the abundance of \ion{He}{} decreased during the quiescent phase compared to the outburst phase of RS Oph. The helium abundance across all epochs was determined by fitting the prominent \ion{He}{I} lines (4026, 4471, 4922, 5016, 5048, 5876, 7065 \AA). Our model clearly shows that the majority of these lines originate from the inner portion of the accretion disc, which is denser than the outer portion. However considerable amount of \ion{He}{} has been generated from the lower density region as well. In addition, during modeling, we observed that a higher density needed to be set to achieve a better fit for these lines, consistent with \citet{2018MNRAS_Zemko}. The generation of helium from higher density regions is also common during active phases of novae (e.g., V1674 Her \citep{2024MNRAS_Habtie, 2024MNRAS_529_917H_Habtie}; RS Oph \citep{2022MNRASPandey}). 

In the initial three epochs (2008, 2010, and 2012), \ion{Fe}{} exhibited subsolar abundances in the accretion disc. However, from the fourth epoch (2014) onward, it showed a considerable increase, appearing overabundant (see Table \ref{tab:T2_results}). The \ion{Fe}{}/\ion{Fe}{$\odot$} ratio in the first and last epochs was obtained as 0.50 and 2.50, respectively. This indicates a considerable enhancement in iron abundance as the nova approaches the upcoming outburst. The possible reason for the enhancement in \ion{Fe}{} could be that the secondary star of RS Oph is an evolved red giant \citet{1996ApJ...456..717_Shore}, and as a result, its outer layers may contain processed material, including heavier elements such as \ion{Fe}{}. The infrared spectrum model of the secondary, conducted by \citet{2008A&A485541_Pavlenko}, also indicates that the metallicity of the secondary in the RS Oph system is enhanced (i.e., [Fe/H] = 0 $\pm$ 0.5). Therefore, the material accreted from the secondary star could potentially be rich in iron. Additionally, the observed enhancement of \ion{Fe}{} abundance might be due to the increase in the accretion rate as the nova system approaches its next outburst.  
The iron abundance for each epoch was determined by fitting specific lines from \ion{Fe}{II} (4233, 4415, 4491, 4584, 4629, 4924, 5018, 5169, 5232, 5276, 5361, and 5538 \AA). These lines predominantly originated from lower-density regions of the disc, with some contributions observed from higher-density areas. 
\ion{Fe}{II} signifies a low ionization stage, suggesting its origin in a zone characterized by low kinetic temperature. During the initial four years from 2008, the abundance of iron was significantly lower than the solar value. This could be due to the limited quantity of accreted matter, resulting in an insufficient amount of iron reaching the disc. The model predicts that the iron abundance surpasses the solar value from 2014 onward, increasing rapidly. This trend aligns with the notion that iron originates from the secondary star. 

\subsubsection{Accretion mass \& rate} \label{amr_rsoph}
We calculate, the accreted mass within the model disc using the following equation, \citep{2001MNRAS-Schwarz}:
\begin{equation}
	M_{\text{disc}} = n(r_0) f(r_0) \int_{R_{\text{in}}}^{R_{\text{out}}} \left(\frac{r}{r_0}\right)^{\alpha+\beta} 2\pi r dr \int_{0}^{h}dz,
\end{equation}
where $n(r_0)$  represents the hydrogen density ($\text{cm}^{-3}$) and $f(r_0)$ stands for the filling factor at the inner radius of the shell ($r_0$). The exponents $\alpha$ and $\beta$ correspond to the power laws. 
The values for density, filling factor, $\alpha$, and $\beta$ are directly adopted from the best-fitting \textsc{cloudy} model parameters (refer to Table \ref{tab:T2_results}).  

\begin{figure}
	\centering
	\includegraphics[scale=0.55]{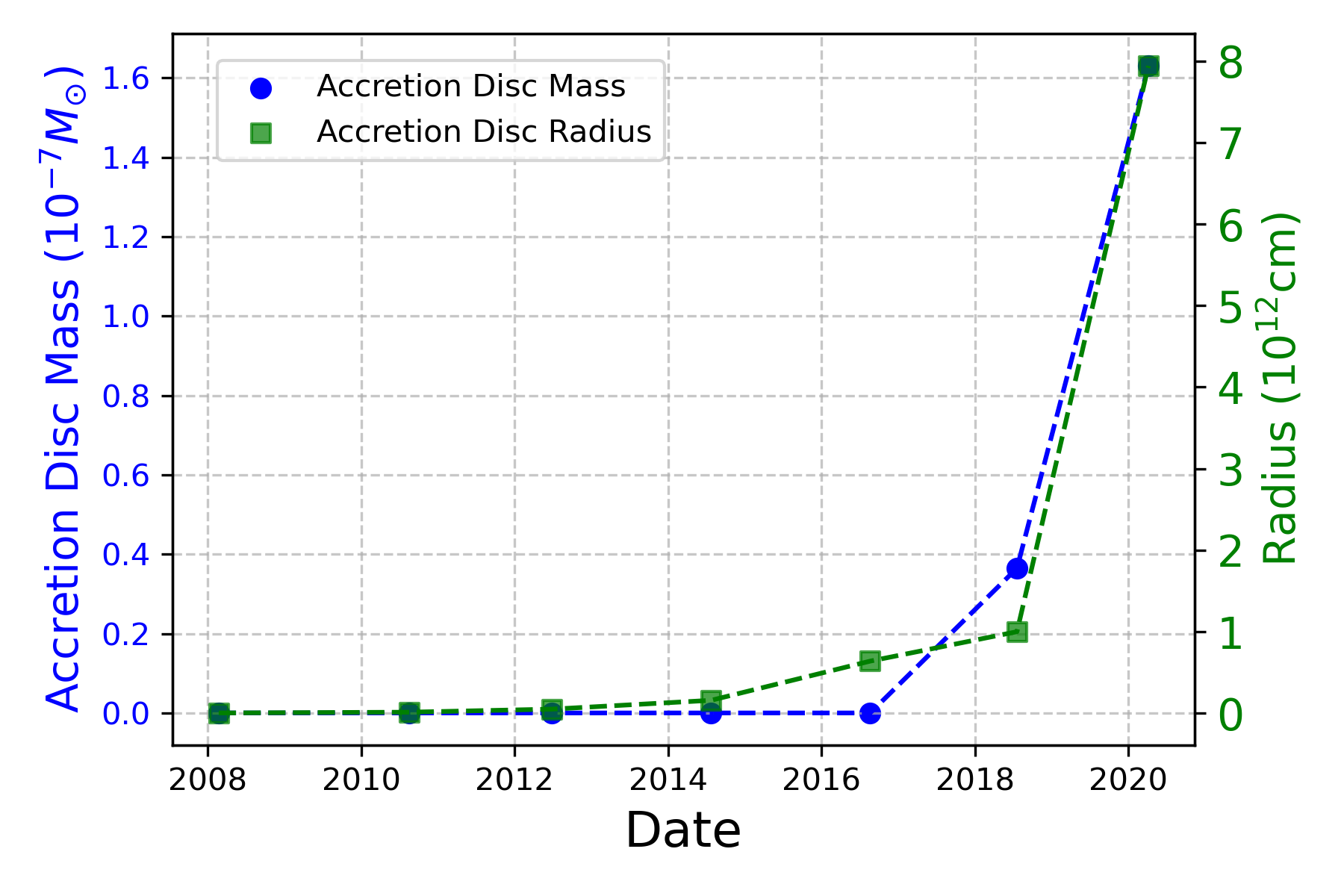}
	\caption{Temporal Evolution of the Accretion Disc's Mass and Radius in RS Oph (2008–2020): This graph showcases the dynamic interplay between accretion disc mass and radius over time, providing a comprehensive view of the system's evolution. The horizontal axis represents the date in years, whereas the vertical axes represent the accretion disc mass (left) and radius (right). 
	}
	\label{fig:accdisc}
\end{figure}

The total accretion disc mass for the two-component model was estimated by multiplying the mass in each density component by their corresponding covering factors and then adding them together. Similarly, for the one-component model, the mass was multiplied by its covering factor. Consequently, the disc masses for epochs 1, 2, 3, 4, 5, 6, and 7 are estimated to be: $~8.94~\times 10^{-16} M_{\odot}$, $~5.39~ \times 10^{-15} M_{\odot}$, $~2.52~ \times 10^{-14} M_{\odot}$, $~4.03~ \times 10^{-14} M_{\odot}$, $8.02 \times 10^{-14} M_{\odot}$, $3.65 \times 10^{-8} M_{\odot}$,  and  $1.63 \times 10^{-7} M_{\odot}$, respectively. In the first epoch, as anticipated, the WD accreted a relatively small mass. The accretion mass showed a gradual increase during Epochs 2, 3, 4, and 5. In these epochs, the accretion mass exhibited a relatively slower growth rate. During Epochs 6 and 7, the accretion disc's mass increased rapidly. Fig. \ref{fig:accdisc} illustrates the growth of the accretion disc over time in terms of both mass and radius. The observed acceleration of accretion rate can be attributed to various factors such as heating of the $L_1$ point by the outer disc \citep{2008A&A_489_699V_Viallet} and/or spiral shocks in the accretion disc \citep{2019MNRAS_483_1080_Pala}. However, these may not be the only factors responsible for the gradual enhancement of accretion rate rather the marginal increase of the system mass may have also a its contribution for the observed acceleration of accretion rate even though it may not be significant.

Counting from the resumption of the accretion disc on the surface of the WD (April 2007) to the last epoch of our model (April 2020), the time span of the accretion disc formation is $\sim$ 13 years. Utilizing the accreted masses estimated from the best-fit parameters of our photoionization model, we calculated a mean accretion rate of  $~\sim~1.25 \times 10^{-8}M_{\odot}$ $\text{yr}^{-1}$. This value is in excellent agreement with previous estimates independently made by \citet{2000ApJ_Hachisu} and \citet{2011ApJ_Nelson}, both of which were $~\sim~1.20 \times 10^{-8}M_{\odot}$ $\text{yr}^{-1}$.

The critical accretion mass necessary for initiating thermonuclear runaway can be estimated by considering the critical pressure in the inner layers of the disc ($10^{15}$ $N~\text{cm}^{-2}$) \citep{1986ApJ_Truran}, as expressed by the formula:
\begin{equation}\label{macc}
	M_{\text{acc}} \approx \dfrac{4\pi R_{\text{WD}}^4}{GM_{\text{WD}}}P_{\text{crit}},
\end{equation}
where $G=6.67\times10^{-11}$ $Nm^2kg^{-2}$, and $R_{\text{WD}}$ and $M_{\text{WD}}$ represent the radius and mass of the WD. Additionally, we determine the WD radius using the \citet{1972ApJ_Nauenberg} mass-radius approximation given in \citet{2005ApJ_Yaron}:
\begin{equation}
	R_{\text{WD}} \approx 1.12\times10^{-2}\left[\left(\dfrac{M_{\text{WD}}}{M_{ch}}\right)^{-2/3}-\left(\dfrac{M_{\text{WD}}}{M_{ch}}\right)^{2/3}\right]^{1/2}R_{\odot}.
\end{equation}

Adopting the WD mass of RS Oph as $M_{WD} = 1.35M_{\odot}$ \citep{2007ApJ_Hachisu}, we obtained a WD radius ($R_{\text{WD}}$) of $\sim$ $1.72~ \times 10^8$ cm. The critical mass required to be accreted onto the WD was then computed using Equation \ref{macc} to be $~M_{\text{acc}}~=~3.07 \times 10^{-7}M_{\odot}$. This is considered as the critical mass of the accreted matter essential for initiating TNR for a WD mass of 1.35$M_{\odot}$.

Our model shows that by July 20, 2018 (epoch 6), the mass of the accreted disc surrounding the white dwarf was approximately $3.65 \times 10^{-8} M_{\odot}$, leaving a deficit of $2.70 \times 10^{-7} M_{\odot}$ to reach the critical mass of $3.07 \times 10^{-7} M_{\odot}$. Since the total mass of the accretion disc is greater than the critical mass required to be accreted onto the white dwarf (WD) to initiate a thermonuclear runaway (TNR), we conclude that only about 12\% of the critical mass had been accreted onto the disc by July 20, 2018 (over a period of approximately 10 years). Consequently, more than 88\% of the total mass of the disc (i.e., $M_{\rm acc} +$ remaining disc mass) had to be accreted in the subsequent three years, leading up to the outburst in August 2021. Similarly, on April 6, 2020, our model estimated an accreted mass of $1.63 \times 10^{-7} M_{\odot}$, leaving a deficit of approximately $1.44 \times 10^{-7} M_{\odot}$ to reach the critical mass. This indicates that only about 53\% of the critical mass had been accreted by April 6, 2020 (over $\sim$13 years), with more than 47\% of the total disc mass (i.e., $M_{\rm acc} +$ remaining disc mass) being accreted over the subsequent 16 months.

 This observation underscores that the accretion rate in the final years is significantly higher compared to earlier periods, driven by the interplay of heating mechanisms and structural dynamics in the accretion disc, which can lead to observable changes in accretion behavior.

\subsubsection{Electron density \& temperature Vs Depth}
Fig. \ref{fig:denvst} in the appendix shows that the electron density and electron temperature are higher in the inner portion of the disk, which is closer to the central star, and then gradually decrease with increasing depth. The distribution appears to follow a certain power-law relation.  From our photo-ionization model we found that the electron temperature on the illuminated face of the accretion disc in all epochs (i.e. at depth = 0 cm or at radius = $7.08 \times 10^8$ cm from the center of the WD) to be 3.47 $\pm$ 0.42 $\times 10^4$K. The electron density in all epochs showed a considerable increment from 3.76 $\times$ $10^9$ to 3.40 $\times$ $10^{10} ~cm^{-3}$.  Fig. \ref{fig:denvst} in the appendix, illustrates the variation of the electron temperature and density with the depth into the accretion disc from the illuminated face to the WD. Both the electron temperature and densities in each epoch showed a gradual decrease in going from the inner radius to the outer radius.

\subsubsection{Lines Volume Emissivity in the disc}\label{LVED}
The volume emissivity of a line describes the energy emitted in a specific spectral line per unit time and per unit volume. From our best-fit model, we have obtained the volume emissivity of selected prominent lines in the disc with respect to depth (the distance between the illuminated face of the disc and a point within the disc). Fig. \ref{fig:ems} in the appendix, shows how the volume emissivity of lines \ion{H}{$\alpha$}, \ion{H}{$\beta$}, \ion{H}{$\gamma$}, \ion{H}{$\delta$}, and \ion{He}{I} 5876 \AA~ varies relative to depth measured from the illuminated face of the disc to a point within the disc. The line volume emissivity in the disc decrease while moving deeper into the disk from the surface of the WD, due to the decrease in temperature and the increase in optical depth \citep{2002apa_book_Frank}, Consequently the photons produced at greater depths are more likely to be absorbed or scattered before they can escape the disk. Therefore the disk becomes more opaque at greater depths, reducing the amount of radiation that can be emitted outward. The volume emissivity has a direct relationship with the square of the electron density and the square root of the electron temperature ($4\pi j_{br}\propto N_e^2T_e^{1/2}$), where $j_{br}$ stands for bremsstrahlung emissivity \citep{2002apa_book_Frank}. This demonstrates that the obtained emissivity is consistent with the $T_e$ and $n_e$ values in appendix Fig. \ref{fig:denvst}.

\begin{landscape}
	\begin{table}
		\centering
		\caption{Observed and best-fit \textsc{cloudy} model line flux values for quiescent phase epochs of RS Oph.}
		\label{tab:T3_chi_square}  
		\begin{tabular}{llccccccccccccccccccccc}
			\hline
			Line&&\multicolumn{3}{c}{Epoch 1 (2008Feb22)}&\multicolumn{3}{c}{Epoch 2 (2010Aug15)}&\multicolumn{3}{c}{Epoch 3 (2012Jun26)}&\multicolumn{3}{c}{Epoch 4 (2014Jul24)}&\multicolumn{3}{c}{Epoch 5 (2016Aug20)}&\multicolumn{3}{c}{Epoch 6 (2018Jul20)}&\multicolumn{3}{c}{Epoch  (2020Apr06)}\\
			\cmidrule(lr){3-5}\cmidrule(lr){6-8}\cmidrule(lr){9-11}\cmidrule(lr){12-14}\cmidrule(lr){15-17}\cmidrule(lr){18-20}\cmidrule(lr){21-23}
			ID&$\lambda$ ({\AA})&mod.&Obs.&$\chi^2$&mod.&Obs.&$\chi^2$&mod.&Obs.&$\chi^2$&mod.&Obs.&  
			$\chi^2$&mod.&Obs.& $\chi^2$&mod.&Obs.&$\chi^2$&mod.&Obs.&$\chi^2$\\
			\hline
			\ion{H}{8}          &3770&0.304&0.271&0.074&-    &-    &-    &-    &-    &-    &-    &-    &-    &-    &    
			-    &-    &-    &-    &-    &-    &-    &   \\
			\ion{H}{$\eta$}     &3835&0.190&0.185&0.002&-    &-    &-    &-    &-    &-    &-    &-    &-    &-    &
			-    &-    &-    &-    &-    &-    &-    &-   \\
			\ion{H}{$\zeta$}    &3889&0.511&0.125&0.048&0.312&0.331&0.031&0.290&0.287&0.001&0.366&0.172&2.239&-    &     
			-    &-    &-    &-    &-    &-    &-    &-   \\
			\ion{H}{$\epsilon$} &3970&0.100&0.188&0.539&0.144&0.248&0.886&0.181&0.441&3.982&0.225&0.336&0.727&0.153& 
			0.066&0.363&-    &-    &-    &-    &-    &-    \\
			\ion{He}{i}         &4026&-    &-    &-    &-    &-    &-    &0.315&0.376&0.218&0.197&0.111&0.439&0.122&  
			0.134&0.007&-    &-    &-    &-    &-    &-    \\
			\ion{H}{$\delta$}   &4101&0.247&0.355&0.812&0.812&0.424&0.005&0.375&0.277&0.575&0.389&0.334&1.885&0.356& 
			0.470&0.878&0.090&0.202&0.027&0.211&0.232&0.086\\
			\ion{Fe}{ii}        &4233&0.091&0.088&0.001&0.079&0.220&1.643&0.076&0.242&1.634&0.082&0.087&0.002&0.104&
			0.116&0.007&1.117&0.073&0.014&0.054&0.095&0.346\\
			\ion{H}{$\gamma$}   &4340&0.885&0.743&1.403&0.738&0.688&0.207&1.083&0.869&1.709&0.861&0.661&2.376&0.737& 
			0.533&1.979&0.438&0.284&1.641&0.430&0.204&10.378\\
			\ion{Fe}{ii}        &4415&0.226&0.110&0.938&0.184&0.154&0.075&0.174&0.412&3.357&0.230&0.303&0.311&0.180& 
			0.082&0.457&0.151&0.040&0.853&0.043&0.030&0.0314\\
			\ion{He}{i}         &4471&-    &-    &-    &-    &-    &-    &-    &-    &-    &-    &-    &-    &0.222& 
			0.339&0.647&0.128&0.037&0.568&0.128&0.037&0.108\\
			\ion{Fe}{ii}        &4491&0.212&0.237&0.042&-    &-    &-    &-    &-    &-    &-    &-    &-    &-    &
			-    &-    &-    &-    &-    &-    &-    &- \\
			\ion{Fe}{ii}        &4584&0.312&0.318&0.003&0.246&0.351&0.923&0.420&0.500&0.376&0.367&0.552&2.016&0.328&   
			0.162&1.313&0.143&0.107&0.087&0.057&0.059&0.001 \\
			\ion{Fe}{ii}        &4629&0.517&0.527&0.006&-    &-    &-    &0.547&0.765&2.821&-    &-    &-    &0.450&   
			0.249&2.104&0.189&0.026&1.841&0.025&0.071&0.429\\
			\ion{H}{$\beta$}    &4861&1.000&1.000&0.000&1.000&1.000&0.000&1.000&1.000&0.000&1.000&1.000&0.000&1.000&  
			1.000&0.000&1.000&1.000&0.000&1.000&1.000&0.000\\
			\ion{He}{i}         &4922&0.398&0.591&2.589&0.252&0.703&0.581&0.280&0.459&1.894&0.469&0.458&0.007&0.162& 
			0.283&0.705&-    &-    &-    &-    &-    &-   \\
			\ion{Fe}{ii}        &4924&-    &-    &-    &-    &-    &-    &-    &-    &-    &-    &-    &-    &-    &   
			-    &-    &0.205&0.111&0.607&0.099&0.099&0.001\\
			\ion{He}{i}         &5016&0.318&0.515&2.682&0.224&0.035&0.029&0.341&0.547&2.512&0.526&0.651&0.911&0.166& 
			0.311&1.008&-    &-    &-    &-    &-    &-    \\
			\ion{Fe}{ii}        &5018&-    &-    &-    &-    &-    &-    &-    &-    &-    &-    &-    &-    &-    &  
			-    &-    &0.156&0.125&0.066&0.128&0.124&0.002\\
			\ion{He}{i}         &5048&0.372&0.349&0.035&-    &-    &-    &-    &-    &-    &-    &-    &-    &0.100& 
			0.272&1.408&-    &-    &-    &-    &-    &-    \\
			\ion{Fe}{ii}        &5169&0.199&0.290&0.572&0.200&3.681&3.042&0.191&0.075&0.801&0.425&0.505&0.385&0.149& 
			0.214&0.201&0.190&0.169&0.031&0.139& 0.082&0.645\\
			\ion{Fe}{ii}        &5232&0.241&0.356&0.915&0.238&0.011&0.013&0.414&0.355&0.208&0.273&0.463&2.151&0.148& 
			0.116&0.049&0.192&0.111&0.449&0.042&0.060&0.062\\
			\ion{Fe}{ii}        &5276&0.364&0.585&3.394&0.267&0.418&1.880&0.237&0.383&1.257&0.419&0.291&0.973&0.213& 
			0.308&0.422&0.244&0.082&1.827&0.076&0.068&0.080\\
			\ion{Fe}{ii}        &5361&0.533&0.785&4.396&0.346&0.268&0.501&0.351&0.468&0.799&0.520&0.616&0.547&-    &   
			-    &-    &0.295&0.072&3.457&0.067&0.061&0.001\\
			\ion{Fe}{ii}        &5538&-    &-    &-    &-    &-    &-    &-    &-    &-    &-    &-    &-    &0.181& 
			0.228&0.105&0.277&0.063&3.183&0.031&0.043&0.065\\
			\ion{He}{i}         &5876&0.509&0.565&0.216&0.519&0.558&0.125&0.499&0.391&0.696&0.734&0.456&4.554&0.506& 
			0.454&0.128&0.315&0.233&0.466&0.243&0.138&3.331\\
			\ion{H}{$\alpha$}   &6563&-    &-    &-    &3.462&3.376&0.610&-    &-    &-    &-    &-    &-    &3.537&   
			4.017&10.940&-    &-    &-    &-    &-    &-    \\
			\ion{He}{i}         &6678&0.599&0.730&1.201&0.281&0.348&0.375&0.219&0.252&0.065&0.294&0.458&1.588&0.146 
			&0.252&0.533&0.075&0.105&0.063&0.073&0.064&0.034\\
			\ion{He}{i}         &7065&0.555&0.682&1.145&0.251&0.325&0.442&0.209&0.264&0.177&0.374&0.452&0.369&0.116&
			0.129&0.009&0.079&0.097&0.022&0.049&0.070&0.092\\
			\hline	
		\end{tabular}\\
	\end{table}
\end{landscape}

\begin{landscape}
	\begin{table}
		\centering
		\caption{Best Fit \textsc{cloudy} Model Parameters for Quiescent Stages of RS Oph: 2006 to 2021 Outburst. }
		\label{tab:T2_results}
		\begin{tabular}{lllllllllr}
			\hline
			\multirow{2}{*}{Parameters}  &  \multicolumn{9}{c}{Values}\\
			\cline{2-10}
			&Epoch 1&Epoch 2&Epoch 3&Epoch 4&Epoch5&\multicolumn{2}{c}{Epoch6}&\multicolumn{2}{c}{Epoch7}\\
			\cmidrule(lr){7-8}\cmidrule(lr){9-10}
			&&&&&&disc$_{in}$&disc$_{out}$&disc$_{in}$&disc$_{out}$\\
			\midrule
			Blackbody temperature ($\times 10^4$K)$^a$          &$1.047^{+0.212}_{-0.047}$&$1.072^{+0.277}_{-0.025}$&$1.096^{+0.025}_{-0.005}$&$1.148^{+0.054}_{-0.026}$&$1.202^{+0.057}_{-0.027}$&$1.698^{+0.297}_{-0.285}$&$1.096^{+0.026}_{-0.024}$&$1.778^{+0.128}_{-0.193}$ &$1.122^{+0.053}_{-0.050}$\\
			Luminosity ($\times 10^{30}$ \ergs)$^a$              &$0.100^{+0.216}_{-0.090}$&$0.159^{+0.093}_{-0.133}$&$0.316^{+0.082}_{-0.216}$&$0.501^{+0.293}_{-0.185}$&$1.000^{+0.585}_{-0.027}$&$3.981^{+6.020}_{-1.470}$&$3.981^{+6.020}_{-1.470}$&$7.940^{+23.68}_{-2.930}$  &$7.940^{+23.68}_{-2.930}$\\
			Hydrogen density ($\times 10^{10}\text{cm}^{-3}$)$^a$&$0.316^{+1.296}_{-0.216}$&$1.000^{+9.000}_{-0.840}$&$3.162^{+1.850}_{-2.162}$&$3.981^{+6.020}_{-2.980}$&$6.309^{+9.540}_{-4.720}$&$10.00^{+316.2}_{-8.400}$&$1.000^{+2.162}_{-0.801}$&$31.62^{+284.6}_{-21.600}$ &$3.162^{+6.840}_{-2.160}$\\
			$\alpha$$^b$	                                     &-2.000&-2.000&-2.000&-2.000&-2.000&-2.000&-2.000 
			&-2.000 &-2.000\\
			Inner radius ($\times 10^{9}\text{cm}$)$^b$	     &0.708&0.708&0.708&0.708&0.708&0.708&31.62&0.708 
			&31.622\\
			Outer radius ($\times 10^{11}\text{cm}$)$^{a\dagger}$     &0.032 &0.126& 0.50&1.584& 6.309&0.316&10.00&0.316 
			&100.0\\
			Filling factor$^b$	                             &0.100&0.100&0.100&0.100&0.100&0.100&0.100&0.100 
			&0.100\\
			$\beta$$^b$	                                     &0.000&0.000&0.000&0.000&0.000 &0.000&0.000&0.000 
			&0.000\\
			Covering factor (AD:BB:SE)$^a$                       &45.5:15.5:39.0 &62.5:15.5:22.0 &63.0:17.5:19.5  
			&59.0:10.0:33.0&65.0:16.0:17.0& 42.0:3.0:8.0 &44.0:3.0:8.0 &43.0:2.0:3.0&49.0:3.0:3.0\\
			\ion{He}{}/\ion{He}{}$_{\sun}$$^{a*}$               &$2.400^{+0.200}_{-0.100}$ &$2.000^{+0.400}_{-0.300}$ & $2.100^{+0.400}_{-0.200}$ &$2.100^{+0.500}_{-0.200}$&$2.100^{+0.600}_{-0.300}$ 
			&$1.100^{+0.200}_{-0.200}$&$1.100^{+0.200}_{-0.200}$&$1.000^{+0.200}_{-0.200}$ &$1.000^{+0.200}_{-0.200}$\\
			\ion{Fe}{}/\ion{Fe}{}$_{\sun}$$^{a*}$               &$0.500^{+0.100}_{-0.100}$&$0.500^{+0.100}_{-0.200}$&$0.800^{+0.400}_{-0.100}$&$1.900^{+0.300}_{-0.400}$&$2.100^{+0.400}_{-0.400}$&$2.400^{+0.300}_{-0.200}$&$2.400^{+0.300}_{-0.200}$&$2.500^{+0.500}_{-0.500}$ &$2.500^{+0.500}_{-0.500}$\\		
			Number of lines &22.000 &17.000 &20.000&18.000&19.000 &\multicolumn{2}{c}{19.000}&19.000&19.000\\
			Number of free parameters &7.000& 7.000& 7.000&7.000&7.000&\multicolumn{2}{c}{7.000}&7.000&6.000\\
			Degrees of freedom &15.000 &10.000 &13.000 &11.000&12.000&\multicolumn{2}{c}{11.000}&12.000&13.000\\
			$\chi_{\text{tot}}^2$ & 21.012&11.365&23.082&21.477&23.261&\multicolumn{2}{c}{15.202}&15.691&15.691 \\
			$\chi_{\text{red}}^2$ &1.401&1.137&1.776&1.953&1.938&\multicolumn{2}{c}{1.382}&1.308&1.207\\
			\hline
		\end{tabular}\\
		{\raggedright Note:  $^a$ and $^b$ refer to the input parameters, which are the free and non-free quantities, respectively. \\
			\hspace{0.6cm} $^{a\dagger}$ refers to the input parameter that is a free parameter, except in the second component of the last epoch. \\
			\hspace{0.6cm} $^{a*}$ Abundances are given on a logarithmic scale of solar number abundance relative to hydrogen, \ion{He}{} = -1.00, and \ion{Fe}{} = -4.55 \citep{2001AIPC59823H_Holweger, 2010ApGrevesse}. We have kept other \\ \hspace{0.9cm} elements either at solar values or turned off any elements that do not appear in the observed spectrum.
			 
			\par}
	\end{table}
	\vspace{2cm} 
\end{landscape}

\section{Conclusions}\label{C}
Our investigation of RS Oph during the quiescent phase between the 2006 and 2021 outbursts employed photoionization-based modeling and spectroscopic analysis to elucidate the accretion disc formation process, encompassing its composition, mass, and dimensions. The key findings are summarized below:
\begin{enumerate}[i.]
	\item The quiescent phase spectra shows low-ionization emission features, including hydrogen, helium, iron, and TiO absorption features.
	\item The high-resolution spectral profiles of \ion{H}{$\alpha$} and \ion{H}{$\beta$} showed a deep absorption at the top, resulting in double-peaked profiles. This is due to the slow and dense wind in the system.
	\item The core of \ion{H}{$\alpha$} showed a considerable shift over time towards the either blue or red edges of the profiles, which is due to the orbital motion of the primary and secondary around their common center of mass. Fluctuations in the accretion disc could be another possible reason for this shift.
	\item The width of the Balmer lines showed a continuous decrease, except for the last spectrum, which was taken approximately five months before the outburst in 2021. While this was mainly due to differences in the resolving power of the telescopes, the slowing down of the remaining ejecta and an increase in the accretion rate may have also contributed.
	\item Using the double peak features observed in the \ion{H}{$\alpha$} and \ion{H}{$\beta$} lines, we have estimated the accretion disc size to be \( R_{AD} = 3.10 \pm 0.04 \times 10^{12} \, \text{cm} \).
	\item Utilizing the \textsc{cloudy} photoionization code, we determined  the temperature of the central ionizing sources in the range of $1.05 - 1.80~\times 10^4$ K and luminosities between $0.10 - 7.90~\times 10^{30}$ \ergs.
	\item The abundance of \ion{He}{} displayed temporal variations, showing an overabundance from 2008 to 2016, returning to solar values by 2020. Meanwhile, \ion{Fe}{} appeared subsolar from 2008 to 2014.
	\item The mean accretion rate, as calculated from the model, is $\sim$ $1.25 \times 10^{-8} M_{\odot}$ yr$^{-1}$. However, it is important to note that this value does not imply uniform accretion dynamics over time. 
	\item The accreted mass in the last 16 months exceeds 47\% of the critical mass, and more than 88\% of the critical mass was accreted in the last three years. This rapid increase in accretion rate within the final years possibly attributed to heating of the L$_1$ point by the outer disc and/or spiral shocks in the accretion disc, influencing the accretion dynamics as the system approaches the critical mass limit.
	\item The critical mass of the accretion disc  is calculated to be  $3.07 \times 10^{-7}M_{\odot}$. 
	\item  From our model we found that the electron temperature ($T_e$) and electron density ($n_e$) to be 3.47 $\pm$ 0.42 $\times 10^4$K and  3.76 $\times$ $10^9$ to 3.40 $\times$ $10^{10} ~cm^{-3}$, respectively. 
\end{enumerate}

\section*{Acknowledgments}
We sincerely thank the anonymous referee for their insightful feedback, which has significantly improved this paper. We thank the S. N. Bose National Centre for Basic Sciences and The World Academy of Science (TWAS) for their funding support. We are grateful to F. Teyssier for coordinating the ARAS Eruptive Stars Section. We also appreciate the ARAS observers for generously sharing their observations with the public. Additionally, we are thankful to David Boyd, Pavol A. Dubovsky, Christian Buil, Joan Guarro Flo, Tim Lester, and Stony Brook for their valuable spectroscopic observations. Additionally, we acknowledge the observers at Stony Brook and ESO for publicly available data. We extend our thanks to Dr. Anindita Mondal for her discussions and email guidance on utilizing \textsc{cloudy} modeling scripts for quiescent stages. GRH acknowledges the support from Debre Berhan University, Debre Berhan, Ethiopia.

\section*{Data Availability}
The paper utilizes spectroscopic data obtained from three sources: the Astronomical Ring for Access to Spectroscopy Database (ARAS Database\footnote{ \url{https://aras-database.github.io/database/novae.html}}; \citet{2019Teyssier}), Stony Brook/SMARTS Atlas of (mostly) Southern Novae (\footnote{\url{http://www.astro.sunysb.edu/fwalter/SMARTS/NovaAtlas/rsoph/rsoph.html}}\citep{2012PASPWalter}), European Southern Observatory (ESO) \footnote{\url{https://www.eso.org/sci/facilities/paranal/decommissioned/isaac/tools/lib.html}} \citep{Pickles_1998PASP}, and Astrosurf Recurrent Nova\footnote{\url{http://astrosurf.com/buil/us/rsoph/rsoph.htm}}. 


\bibliographystyle{mnras}
\bibliography{Rs_Oph_Submited_new} 



\appendix
\section{Electron density and Temperature in the accretion disc}
\begin{figure*}
	\centering
	\includegraphics[scale=0.9]{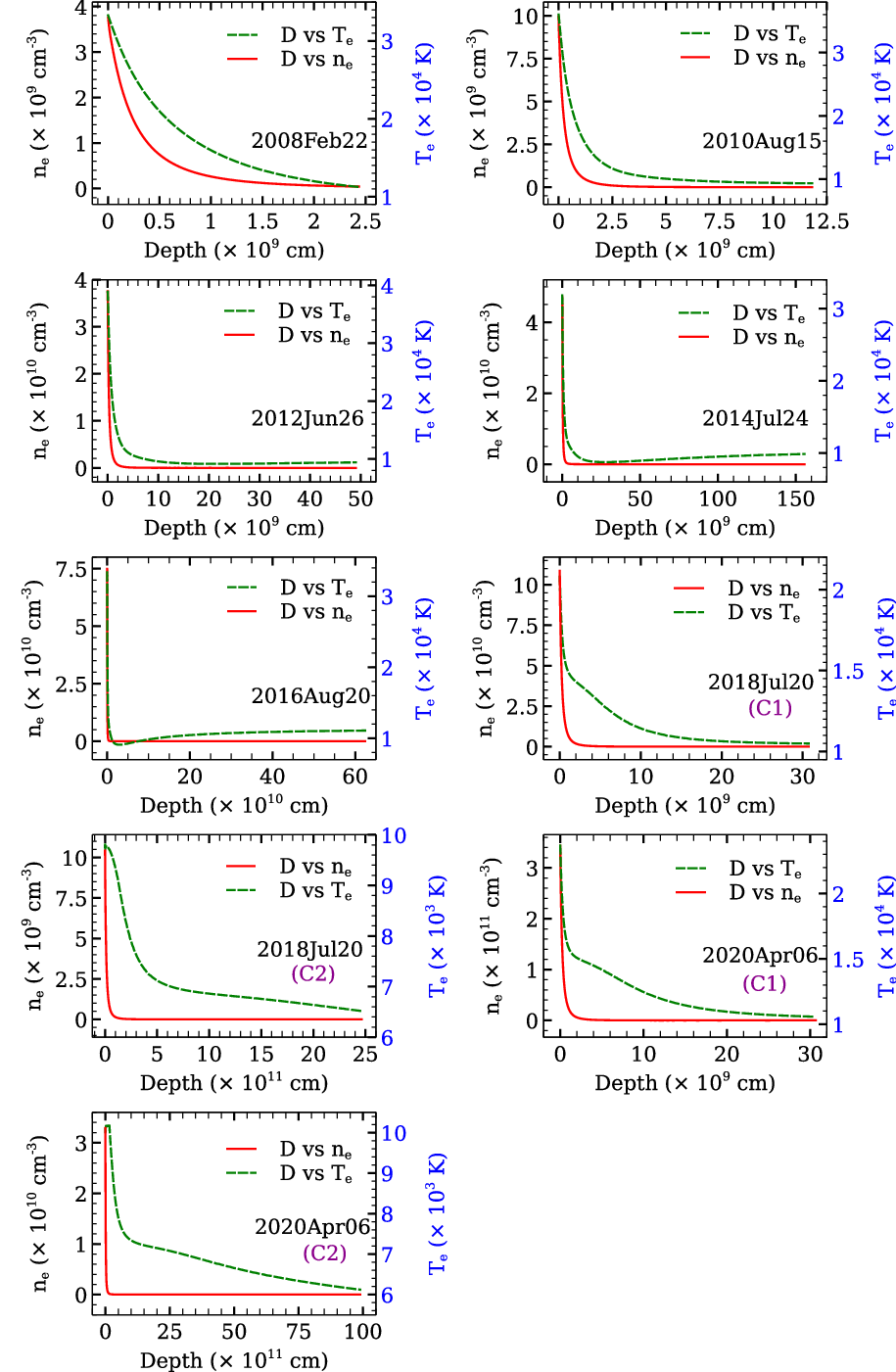}
	\caption{ Variation of electron density ($n_e$) and electron temperature ($T_e$) with respect to depth (D) inside the accretion disc from the illuminating face of the disc. The red solid line represents $D$ vs. $n_e$, and the green dashed line represents $D$ vs. $T_e$. The C1 and C2 labels in the last two epochs indicate components 1 and 2, representing the inner and outer portions of the disc, respectively.
	}
	\label{fig:denvst}
\end{figure*}

Fig. \ref{fig:denvst} illustrates that the maximum value of both the electron density ($n_e$) and electron temperature ($T_e$) lies at the illuminating face of the accretion disc which is facing the surface of the WD. While going deeper into the disc both the $n_e$ and $T_e$ decreases . 

\section{Emissivity vs Depth}
\begin{figure*}
	\centering
	\includegraphics[scale=0.8]{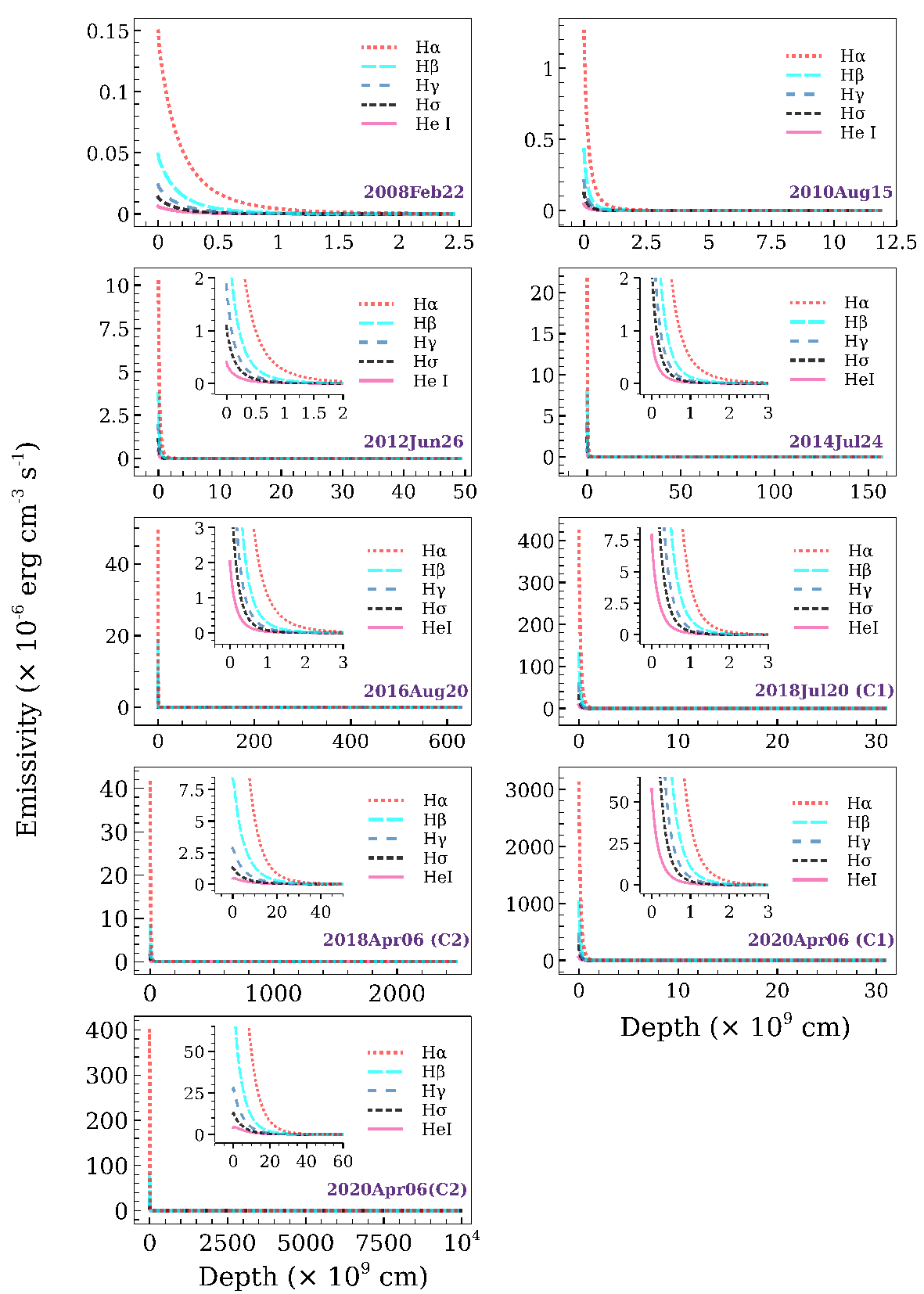}
	\caption{ Volume emissivity of \ion{H}{$\alpha$}, \ion{H}{$\beta$}, \ion{H}{$\gamma$}, \ion{H}{$\delta$}, and  \ion{He}{I} 5876 \AA~ in disc with respect to depth (D). The C1 and C2 labels in the last two epochs indicate components 1 and 2, representing the inner and outer portions of the disc, respectively.
	}
	\label{fig:ems}
\end{figure*}

Fig. \ref{fig:ems} shows that the highest probability of emission for the most prominent recombination lines, such as \ion{H}{$\alpha$}, \ion{H}{$\beta$}, \ion{H}{$\gamma$}, \ion{H}{$\delta$}, and \ion{He}{I} 5876 \AA~, is from the illuminated face of the disk. The volume emissivity decreases exponentially with increasing depth.

\bsp	
\label{lastpage}
\end{document}